\documentclass[aps,pra,reprint,superscriptaddress]{revtex4-2}
\usepackage{graphicx}
\usepackage{amsmath}
\usepackage{xcolor}
\usepackage{braket}
\usepackage{amssymb}
\def\Lie#1{\hbox{\sf #1}}

\def\iso#1{\Lie{iso($#1$)}}

\begin{document}
\preprint{}
\title{Mathieu-Bragg photonic lattices}
\author{I. Ramos-Prieto}
\email[e-mail:~]{iranrp123@gmail.com}
\affiliation{Instituto de Ciencias Físicas, Universidad Nacional Autónoma de México\\ Apdo. Postal 48-3, Cuernavaca, Morelos 62251, Mexico}
\author{K. Uriostegui}
\affiliation{Instituto de Ciencias Físicas, Universidad Nacional Autónoma de México\\ Apdo. Postal 48-3, Cuernavaca, Morelos 62251, Mexico}
\author{J. Récamier}
\affiliation{Instituto de Ciencias Físicas, Universidad Nacional Autónoma de México\\ Apdo. Postal 48-3, Cuernavaca, Morelos 62251, Mexico}
\author{F. Soto-Eguibar}
\affiliation{Instituto Nacional de Astrofísica Óptica y Electrónica, Calle Luis Enrique Erro No. 1, Santa María Tonantzintla, Pue., 72840, Mexico}
\author{H. M. Moya-Cessa}
\affiliation{Instituto Nacional de Astrofísica Óptica y Electrónica, Calle Luis Enrique Erro No. 1, Santa María Tonantzintla, Pue., 72840, Mexico}
\date{\today}
\begin{abstract}
We show that Bragg diffraction may be modeled by classical light propagation in photonic lattices  having a square power law for the refraction index coefficient. The dynamics is shown to be fully integrable and therefore described in closed form. We examine the trajectories of classical light propagating in such structures.
\end{abstract}
\maketitle
\textit{Introduction.}
The propagation of light in photonic lattices has attracted a lot of interest in recent years. It is possible to engineer photonic lattice configurations such that the flow of light may be predicted, giving rise to possible applications. The discrete coupling or tunneling process between periodically arranged potential wells is a fundamental topic that has been extensively investigated \cite{Garanovich_2009}. In optics, arrays of weakly coupled waveguides and resonators are excellent examples of such systems, where the coupling dynamics can be directly observed and investigated \cite{Christodoulides_2003,Yariv,Snyder_1984,Jones_1965}. The similarities between wave optics and quantum mechanics lead to judiciously established analogies in photonic structures \cite{Longhi_2009a,Leija_2010}, since the electric field and the wave function obey the same equation: the paraxial equation and Schr\"odinger equation. For instance, and of particular interest, when the refractive index is constant but different in each waveguide, the dynamics of the light in the photonic lattice exhibits Bloch oscillations \cite{Peschel_1998}, and even more, as long as the optical potential is periodic (complex or real), the dynamics of an atom within a crystalline lattice can be mapped to these types of evanescent structures \cite{Keller_1997,Longhi_2009b,Longhi_2010}. Significantly, when a two-level atom interacts with two counterpropagating light fields (standing field), that is, a periodic optical potential, it is well-known that atomic Bragg diffraction obeys systems of differential equations \cite{Peter_1987,Muller_2008,Giese_2013,Giese_2015,Hartmann_2020} resembling the ones obtained in classical light propagation in photonic lattices \cite{Perez-Leija_2013,Keil_2012,Rodriguez_2013}; in fact, although they are time dependent, sets of unitary (similarity) transformations (see below) may lead to differential equations commonly obtained for light propagating in in-homogeneous media \cite{Jones_1965} making both systems analogous.

\begin{figure}
\begin{center}
\includegraphics{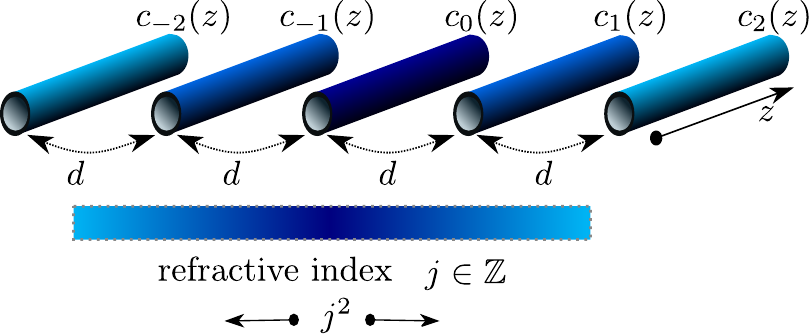}
\caption{Schematic representation of an array of evanescently coupled waveguides. The separation distance is the same ($d$), but $j^2$ corresponds to the propagation constant of the waveguide that obeys a transverse quadratic refractive index. The color of each waveguide specifies a different refractive index around the central waveguide.}
\label{Fig_1}
\end{center}
\end{figure}

In this Letter, we show that an infinite array having a square  law distribution for the transversal refraction index can be a platform to emulate the Bragg diffraction processes. To increase the index of refraction in a lattice, as the one shown in Fig.~\ref{Fig_1}, may be a difficult task, as the coupling has to grow quadratically; therefore, ways of simulating such waveguide array may be of interest. The purpose of the present contribution is to introduce a photonic lattice where light propagation models Bragg diffraction; as the system of differential equations leads us to the Mathieu equation, we name such lattice a Mathieu-Bragg photonic lattice.

\textit{Operator approach to single Bragg diffraction processes.}
In single Bragg diffraction processes, it is considered that a two-level atom exchanges energy and momentum with two counter-propagating light fields of frequencies $\omega_b$ and $\omega_a$. In particular, the first-order diffraction occurs when the laser detuning coincides with the kinetic energy gained via momentum transfer, i.e. $\Delta\omega:=\omega_b-\omega_a=k^{2}/2:=\omega_k$. It is also assumed that the internal states of the atom do not change after the Bragg processes; therefore, if the atom is in its ground state $\ket{g}$, the excited state $\ket{e}$ can be eliminated adiabatically from the dynamics of the system, and the following system of coupled differential equations can be obtained \cite{Giese_2013,Giese_2015,Hartmann_2020}
\begin{align}\label{Eq:g}
\mathrm{i}\frac{dg_j}{dt} =& -\Omega\left[ g_{j-1}e^{\mathrm{i}\omega_D t} e^{\mathrm{i}2\omega_k(j-1)t} +g_{j+1}e^{-\mathrm{i}\omega_D t}e^{-\mathrm{i}2\omega_k j t}\right],
\nonumber \\ &
j\in\mathbb{Z},
\end{align}
where  $g_j(t)$ is the probability amplitude for the ground state in the momentum representation, labeled by index $j$, $2\Omega$ is called the Bragg-Rabi frequency and it determines the coupling strength, $\omega_D$ is the Doppler shift frequency analogous to a detuning for a momentum distribution around a resonance momentum $p$. As a consequence, the resonant condition around this point is defined as: $\Delta\omega=\left[(p+k)^2-p^2\right]/2=k^2/2+pk:=\omega_k+\omega_D$.

On the other hand, to establish the Hamiltonian $\hat{\mathcal{H}}$ associated to the previous set of coupled differential equations, let us define operators and state vectors as
\begin{equation}
\begin{split}
\hat{N}&=\sum_{j=-\infty}^{\infty}j\ket{j}\bra{j},\\
\hat{V}&=\sum_{j=-\infty}^{\infty}\ket{j}\bra{j+1},
\end{split}
\quad
\begin{split}
\hat{V}^\dagger&=\sum_{j=-\infty}^{\infty}\ket{j+1}\bra{j},\\
\ket{\psi(t)}&=\sum_{j=-\infty}^{\infty}g_{j}(t)\ket{j},
\end{split}
\end{equation}
where $g_j(t):=\braket{j|\psi(t)}$. The set of operators $\left\lbrace \hat{N}, \hat{V}, \hat{V}^\dagger \right\rbrace$ creates a Lie algebra $\iso{1,1}$ \cite{Vilenkin_1991}, and obeys the commutation relations
\begin{equation}\label{RC}
    \left[\hat{N},\hat{V}\right]=-\hat{V},\quad[\hat{N},\hat{V}^\dagger]=\hat{V}^\dagger,\quad[\hat{V},\hat{V}^\dagger]=0.
\end{equation}

Under such conditions, it is straightforward to write the differential system (\ref{Eq:g}) as
\begin{align}\label{Eq:psi}
-\mathrm{i}\frac{\partial\ket{\psi(t)}}{\partial t}
=\Omega\bigg[\hat{V}&e^{-\mathrm{i} 2\omega_k\left(\hat{N}-1\right)t}e^{-\mathrm{i}\omega_Dt}
\nonumber \\&
+e^{\mathrm{i}2\omega_k\left(\hat{N}-1\right)t}e^{\mathrm{i} \omega_Dt}\hat{V}^\dagger\bigg] \ket{\psi(t)}.
\end{align}
To remove the time-dependent coefficients, we make the time-dependent unitary transformation $\ket{\psi(t)}=e^{-\mathrm{i}\delta t \hat{N}^2}e^{-\mathrm{i}\eta t\hat{N}} \ket{\phi(t)}$, such that
\begin{equation}\label{0040}
\begin{split}
-\mathrm{i}\frac{\partial \ket{\phi(t)}}{\partial t}&=\bigg\{\delta\hat{N}^2+\eta\hat{N}\\
&+\Omega\bigg[\hat{V}e^{-\mathrm{i}2\hat{N}(\delta+\omega_k)t}e^{\mathrm{i}(\delta-\eta+2\omega_k-\omega_D)t}
\\&+e^{\mathrm{i}2\hat{N}(\delta+\omega_k)t}e^{-\mathrm{i}(\delta-\eta+2\omega_k-\omega_D)t}\hat{V}^\dagger\bigg]\bigg\} \ket{\phi(t)},
\end{split}
\end{equation}
where we have used the relations
\begin{equation}\label{FNV}
\hat{V}f(\hat{N}-1)=f(\hat{N})\hat{V}, \qquad
\hat{V}^\dagger f(\hat{N})=f(\hat{N}-1)\hat{V}^\dagger,
\end{equation}
which are a consequence of the commutation relations \eqref{RC}. By setting $\delta=-\omega_k$ and $\eta=\omega_k-\omega_D$, the time dependent terms are eliminated from \eqref{0040}; thus, we may write the Hamiltonian for the Schrödinger equation (\ref{0040}) as
\begin{equation}\label{Hamiltonian}
\hat{H}=\delta\hat{N}^2+\eta\hat{N}+\Omega(\hat{V}+\hat{V}^\dagger).
\end{equation}
Notice that this Hamiltonian contains the term $\eta\hat{N}$. To cancel this term, we apply the time-independent transformation defined by $\ket{\phi}=\hat{V}^{l}\ket{\psi}$ to the Hamiltonian $\hat{H}$, namely $\hat{\mathcal{H}}=\hat{V}^{\dagger l}\hat{H}\hat{V}^{l}$, and we get
\begin{equation}
    \mathcal{\hat{H}}=\delta\left(\hat{N}-l\right)^2+\eta\left(\hat{N}-l\right)+\Omega\left(\hat{V}+\hat{V}^\dagger\right).
\end{equation}
For the sake simplicity, we choose $l=\eta/2\delta$ and we obtain
\begin{equation}\label{H_F}
    \mathcal{\hat{H}}=\delta\hat{N}^2+\Omega\left(\hat{V}+\hat{V}^\dagger\right),
\end{equation}
where $l$ must be an integer. The condition that $l$ must be an integer can be relaxed in order to find a suitable choice of parameters $\omega_k$ and $\omega_D$; in the case that $l$ is not an integer, we still can do the transformation because $[\hat{V}^k,\hat{N}]=k\hat{V}^k$ for $k$ integer, and as
\begin{equation}
\begin{split}
   [f(\hat{V}),\hat{N}]&=\sum\frac{f^{k}(0)}{k!} [\hat{V}^k,\hat{N}]\\&=\sum \frac{f^{k}(0)}{k!}k\hat{V}^k =f'(\hat{V})\hat{V},
\end{split}
\end{equation}
we have
\begin{equation}
    \hat{V}^{l}\hat{N}\hat{V}^{\dagger l}=\hat{N}\hat{V}^{l}\hat{V}^{\dagger l}+l \hat{V}^{l}\hat{V}^{\dagger l}=\hat{N}+l.
\end{equation}
As we have already seen, the set $\left\{\hat{N},\hat{V},\hat{V}^\dagger\right\}$ is a realization of the algebra  $\iso{1,1}$, however, the Hamiltonian $\hat{\mathcal{H}}$ contains the term $\hat{N}^2$ which makes a closed solution impossible, at least using the properties of the algebra $\iso{1,1}$. In fact, below, we propose a Bloch-Floquet-type solution \cite{Longhi_2010} that allows us to recognize the eigenstates of $\hat{\mathcal{H}}$. Alternatively, in the Raman-Nath regime \cite{Muller_2008} the term kinetic energy  ($\hat{N}^2$) is negligible, so the system will evolve according to the generating function $\exp\left[-\mathrm{i}t\Omega(\hat{V}+\hat{V}^\dagger)\right]=\sum_{k}J_{k}(-2\Omega t)\left(\mathrm{i}\hat{V}^{\dagger}\right)^k$ \cite{Moya_Book_2011}, where $J_k(x)$ are the Bessel functions of the first kind.

The previous algebraic derivation of the Bragg platform, represented by the Hamiltonian (\ref{H_F}), is the first contribution of this article. In the following, we will focus on a waveguide array that precisely gives this kind of interaction, so we may say that those two systems are completely equivalent.

\textit{Mathieu-Bragg waveguide array.} According to the coupled mode theory, the propagation of an optical field in a waveguide array, with nearest-neighbor evanescent coupling, is governed by the set of coupled differential equations \cite{Christodoulides_2003},
\begin{equation}\label{eqc}
\mathrm{i}\frac{dc_{j}(z)}{dz}=j^2c_{j}(z)+q\left[c_{j-1}(z)+c_{j+1}(z)\right],\quad j\in\mathbb{Z},
\end{equation}
where  $z$ represents the propagation distance, and $c_{j}(z)$ is the optical field amplitude at the $j$-th waveguide. We have assumed that each waveguide has the same separation distance and that the coupling strength depends on the parameter $q$, while the refractive index is a constant of propagation but grows quadratically in the transverse direction (see Fig.~\ref{Fig_1}). To establish a one-to-one correspondence between the waveguide array \eqref{eqc}, and the single Bragg diffraction processes, defined by \eqref{H_F}, we choose $q=\Omega/\delta$ and $z=\delta t$, obtaining
\begin{equation}\label{Eq:Full}
    \mathrm{i}\frac{\partial\ket{\Psi(z)}}{\partial z}=\left[\hat{N}^2+q\left(\hat{V}+\hat{V}^\dagger\right)\right]\ket{\Psi(z)},
\end{equation}
where $\ket{\Psi(z)}=\sum_{j}c_{j}(z)\ket{j}$. The Schrödinger-like equation above shows that the Bragg diffraction process and our waveguide array are isomorphic, and this result represents the second main contribution of this work. For example, it is known that the ratio $q=\Omega/\delta$ determines the diffraction regimes: Raman-Nath regime and deep-Bragg regime \cite{Muller_2008,Giese_2013}; in the photonic context, this corresponds to a ballistic behavior when $q\gg1$ and to a localized behavior when $q<1$.

\begin{figure}
\begin{center}
\includegraphics{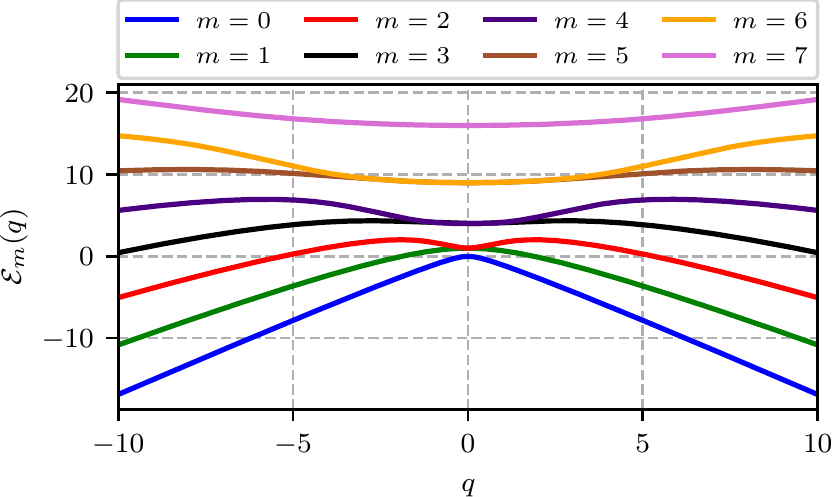}
\caption{(Color online) Eigenvalues or characteristic values $\mathcal{E}_{m}(q)$ as function of $q$ for $m\in\mathbb{N}$. When $q\rightarrow0$ then $\mathcal{E}_{m}(q)\rightarrow\sqrt{m}$, otherwise the characteristic curves~$\mathcal{E}_m(q)$~are divided into two regions of stability and instability inherited from angular Mathieu functions.}
\label{Fig_2}
\end{center}
\end{figure}

In order to explore the diffraction of light in these types of arrays, we consider that the field amplitudes evolve in a way analogous to the matrix elements of the evolution operator \eqref{Eq:Full} $c_n(z;j)=\braket{n|e^{-\mathrm{i}z\left[\hat{N}^2+q(\hat{V}+\hat{V}^\dagger)\right]}|j}$,  solving the distance (or time) independent Schr\"odinger equation is essential to find the amplitude of the optical field at channel $n$ upon excitation of site $j$. Thus, we consider
\begin{equation}\label{Hm}
    \left[\hat{N}^2+q\left(\hat{V}+\hat{V}^\dagger\right)\right]\ket{m;\text{cse}}=\mathcal{E}_{m}(q)\ket{m;\text{cse}}
\end{equation}
with $m\in\mathbb{N}$, and where $\mathcal{E}_{m}(q)$ are the eigenvalues  or  characteristic values to  be  determined, they are parameterized by $q$, and the $\text{cse}$ notation in the ket will be clarified below. To solve \eqref{Hm}, we propose the Bloch-Floquet–type solutions \cite{Longhi_2010}
\begin{equation}\label{Eq:emathieu}
    \ket{m;\text{cse}}=\sum_{j=-\infty}^{\infty}\mathcal{A}_{j}^{(m)}(q)\ket{j},
\end{equation}
where $\mathcal{A}_{j}^{(m)}(q)$ denote $q$-coefficients to be determined; substituting \eqref{Eq:emathieu}  into \eqref{Hm}, we obtain the following recurrence relation
\begin{equation}\label{Eq:Aj}
\begin{split}
    &\left[j^2-\mathcal{E}_{m}(q)\right]\mathcal{A}_{j}^{(m)}(q)+q\left[\mathcal{A}_{j-1}^{(m)}(q)+\mathcal{A}_{j+1}^{(m)}(q)\right]=0,\\& j\in\mathbb{Z}.
\end{split}
\end{equation}
These equations determine the eigenvalues $\mathcal{E}_m(q)$ and the coefficients $\mathcal{A}_{j}^{(m)}(q)$ given the  parameter $q$; it is important to note that each eigenvector, defined by the matrix elements $\mathcal{A}^{(m)}(q):=\lbrace\mathcal{A}_{j}^{(m)}(q),\;j\in\mathbb{Z}\rbrace$, corresponds to the eigenvalue $\mathcal{E}_m(q)$ of the Hamiltonian $\hat{N}^2+q(\hat{V}+\hat{V}^\dagger)$. We now turn our attention to the recurrence relation, since it is known that \eqref{Eq:Aj} determines the coefficients and characteristic values of the angular Mathieu functions, such that the Fourier series of a Floquet solution is \cite{McLachlan_1951,Abramowitz,Ley_2009}
\begin{equation}\label{cse}
    \mathrm{cse}_{m}(x;q)=\sum_{j=-\infty}^{\infty}\mathcal{A}_{j}^{(m)}(q)e^{\mathrm{i}jx}.
\end{equation}
This complex-form solution allows us to recognize the Fourier series form of the elliptic functions $\mathrm{ce}_{2m}(x;q)$ and $\mathrm{se}_{2m}(x;q)$ as follows
\begin{equation}
\begin{split}
	\text{cse}_{m}(x,q) &= \sum_{j=-\infty}^{\infty} \mathcal{A}_{j}^{(m)}(q)e^{\mathrm{i}jx},\\
	&= \mathcal{A}_{0}^{(m)}(q)\\&+\sum_{j=1}^{\infty} \mathcal{A}_{j}^{(m)}(q)e^{\mathrm{i} jx} +\sum_{j=1}^{\infty} \mathcal{A}_{-j}^{(m)}(q)e^{-\mathrm{i} jx}.
\end{split}
\end{equation}
The coefficients $\mathcal{A}_{j}^{(m)}(q)$ have the parities
\begin{equation}
	\begin{cases}
	\mathcal{A}_{j}^{(m)}(q) = \mathcal{A}_{-j}^{(m)}(q)\quad&\text{if}\quad m \quad \text{even},\\
	\mathcal{A}_{j}^{(m)}(q) = -\mathcal{A}_{-j}^{(m)}(q)\quad&\text{if}\quad m \quad\text{odd},
	\end{cases}
\end{equation}
therefore,
\begin{equation}
	\text{cse}_{m}(x,q) =
	\begin{cases}
	\mathcal{A}_{0}^{(2m)}(q) + 2 \sum\limits_{j=1}^{\infty} \mathcal{A}_{j}^{(2m)}(q)\cos(jx), \\
	\mathrm{i}2\sum\limits_{j=1}^{\infty} \mathcal{A}_{j}^{(2m+1)}(q)\sin(jx),
	\end{cases}
\end{equation}
then, by identifying
\begin{equation}
\begin{split}
\mathcal{A}_{0}^{(2m)}(q) = \sqrt{2}A_{0}^{(2m)}(4q),&\quad
\mathcal{A}_{j}^{(2m)}(q)= \textstyle{\frac{1}{\sqrt{2}}} A_{2j}^{(2m)}(4q), \\
\mathcal{A}_{j}^{(2m+1)}(q)&= \textstyle{\frac{1}{\sqrt{2}}}B_{2j+2}^{(2m+2)}(4q),
\end{split}
\end{equation}
where $A_{j}^{(m)}(q)$ and $B_{j}^{(m)}(q)$ are the angular Mathieu coefficients \cite{Abramowitz}. Finally, we have two cases
\begin{equation}\label{cese}
    \text{cse}_{m}(x,q) = \sqrt{2} \begin{cases}
    \text{ce}_{2r}(\textstyle{\frac{x}{2}},4q), \quad & \text{if} \ m \ \text{even}, \\
   \mathrm{i}\,\text{se}_{2r+2}(\textstyle{\frac{x}{2}},4q), \quad & \text{if} \ m \ \text{odd},
    \end{cases}
\end{equation}
with $r \in \mathbb{N}$. The periodicity of the functions $\text{ce}_{2r}(x,q)$ and $\text{se}_{2r+2}(x,q)$ is $\pi$, hence $\text{cse}_{m}(x,q)$ obtains periodicity $2\pi$. This justifies the $\text{cse}$ prefix because it includes both elliptical functions. Normalization relations are inherited from these functions; i.e., $\braket{m;\text{cse}|m;\text{cse}}=1$. Whereupon, the characteristic values are
\begin{equation}\label{valcar}
    \mathcal{E}_{m}(q) =\frac{1}{4} \begin{cases}
    a_{2r}(4q), \quad & \text{if} \ m \ \text{even}, \\
   b_{2r+2}(4q), \quad & \text{if} \ m \ \text{odd},
    \end{cases}
\end{equation}
where $a_{2r}(q)$ and $b_{2r+2}(q)$ are the characteristic values of angular Mathieu functions \cite{Abramowitz} (see Fig.~\ref{Fig_2}).
Once the eigenfunctions (\ref{cese}) have been recognized, and especially the characteristic value and the angular Mathieu coefficients, $\mathcal{E}_{m}(q)$ and $\mathcal{A}_{j}^{(m)}(q)$ respectively,  we must recognize \eqref{Hm} as the angular Mathieu equation with $\braket{x|m;\text{cse}}=\Phi_m(x;q)$, such that
\begin{equation}\label{Eq:MathieuPSI}
\left[-\frac{d^2}{dx^2}+q\left(e^{\mathrm{i}x}+e^{-\mathrm{i}x}\right)\right]\Phi_m(x;q)=\mathcal{E}_m(q)\Phi_{m}(x;q).
\end{equation}
with $\hat{N}=-\mathrm{i}\frac{\partial}{\partial x}$, $\hat{V}=e^{-\mathrm{i}x}$, and $\hat{V}^\dagger=e^{\mathrm{i}x}$, which in turn define the Schrödinger equation for a one-dimensional potential. It is necessary to remark that this type of periodic potentials with non-zero real and imaginary parts exhibit parity-time or $\mathcal{PT}$-symmetry \cite{Keller_1997,Bender1999,Makris_2008,Longhi_2009b,Longhi_2010}. Therefore, it is important to point out, for future approaches, that the operators $\hat{V}$ and $\hat{V}^\dagger$ define the periodic potential in the momentum domain and a suitable combination of them can be used to obtain complex potentials.

By direct integration of \eqref{Eq:Full}, using \eqref{Hm} and \eqref{Eq:emathieu} and the completeness relation $\sum_{m}\ket{m;\text{cse}}\bra{m;\text{cse}}=1$, we can obtain the optical field amplitude at waveguide $n$ after a propagation distance $z$ upon excitation of site $j$, so that
\begin{equation}
\begin{split}
c_{n}(z;j)=&\braket{n|e^{-\mathrm{i}z\left[\hat{N}^2+q(\hat{V}+\hat{V}^\dagger)\right]}|j},\\
=&\sum_{m=0}^{\infty}\braket{n|e^{-\mathrm{i}z\left[\hat{N}^2+q(\hat{V}+\hat{V}^\dagger)\right]}|m;\text{cse}}\braket{m;\text{cse}|j},\\
=&\sum_{m=0}^{\infty}e^{-\mathrm{i}z\mathcal{E}_{m}(q)}\braket{n|m;\text{cse}}\braket{m;\text{cse}|j},\\
=&\sum_{m=0}^{\infty}\mathcal{A}_{n}^{(m)}(q)\mathcal{A}_{j}^{(m)}(q)e^{-\mathrm{i}z\mathcal{E}_{m}(q)}
\end{split}
\end{equation}
where we have used the fact that $\braket{m;\text{cse}|n}=\mathcal{A}_{n}^{(m)}(q)$. In general, when the initial condition is represented by a field distribution, namely $\ket{\Psi(0)}=\sum_{j}c_j(0)\ket{j}$, the impulse response in these initial conditions is given by
\begin{equation}
    c_{n}(z)=\sum_{j=-\infty}^{\infty}\sum_{m=0}^{\infty}c_{j}(0)\mathcal{A}_{n}^{(m)}(q)\mathcal{A}_{j}^{(m)}(q)e^{-\mathrm{i}z\mathcal{E}_{m}(q)}.
\end{equation}
where $c_j(0)$ is the input amplitude in the $j$-th waveguide.

\begin{figure}
\begin{center}
\includegraphics{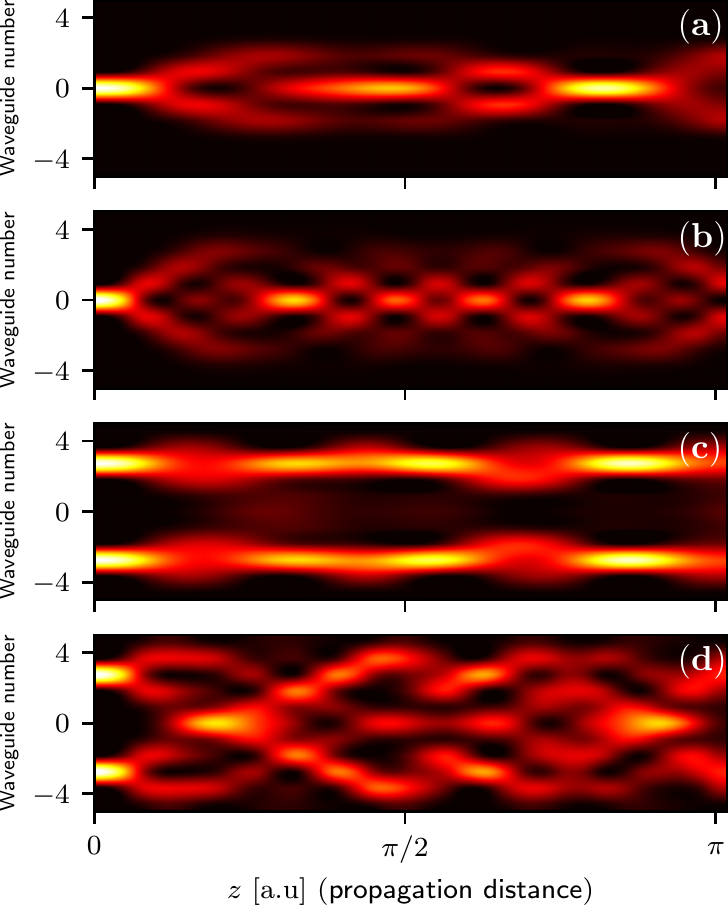}
\caption{(a) Intensity evolution when the central site is excited, (c) and when sites $j = 3$ and $j = -3$ are initially launched at $z=0$. with $q=2$. (b)-(d) The same initial condition as in the previous cases but setting $q = 5$, respectively.}
\label{Fig_3}
\end{center}
\end{figure}

To illustrate the dynamic behavior of light in the Mathieu-Bragg photonic lattice when one or more sites are excited, in Fig.~\ref{Fig_3} we show the intensity evolution of light. In (a)-(b) the central waveguide is excited, while in (c)-(d) two sites are excited symmetrically to the central waveguide, with $q=2,4$ respectively. As we can see, the diffraction of light becomes localized as $q\rightarrow1$. To compute the coefficients $\mathcal{A}_j^{(m)}(q)$ and the characteristic values $\mathcal{E}_{m(q)}$, we use the matrix method established in \cite{Ley_2009}, where the determinant of the tridiagonal matrix associated with equation \eqref{Hm} first determines the characteristic values $\mathcal{E}_{m}(q)$ and in turn one can obtain the coefficients  $\mathcal{A}_{j}^{(m)}(q)$, parameterized by the value $q$.

\textit{Conclusions.} As we already mentioned, this article presents two main contributions. First, the algebraic derivation of the Hamiltonian \eqref{H_F} that describes the Bragg diffraction, and second, the establishment of a waveguide platform to emulate the Bragg processes. We show that this platform is closely related to periodic potentials, and at the same time, with the angular Mathieu equation. This allows us to export all the properties of the Mathieu solutions to the solution of the Schrödinger equation in the momentum domain.

\section*{Acknowledgments}
One of us H.M.M.C. acknowledges useful discussions with Dr. A. Perez-Leija. J.R. and I.R.P. acknowledge partial support from Direcci\'on General de Asuntos del Personal Acad\'emico, Universidad Nacional Aut\'onoma de M\'exico (DGAPA UNAM) through project PAPIIT IN 1111119, and I.R.P acknowledges postdoctoral support from DGAPA UNAM. K.U. acknowledge doctoral fellowship support from CONACyT-México.
%
\end{document}